\documentclass[10pt,letterpaper,twocolumn,english]{article}
\pdfoutput=1  
\usepackage{ol2}
\usepackage{hyperref}
\usepackage[T1]{fontenc}
\usepackage[latin9]{inputenc}
\usepackage{babel}
\usepackage{amsmath}
\usepackage{amssymb}
\usepackage{graphicx}
\usepackage{esint}
\usepackage{bbold}
\usepackage{breakurl}
\usepackage{siunitx}

\newcommand{\mycite}[1]{\cite{#1}}

\usepackage{graphicx,color}

\makeatletter

\DeclareFontEncoding{LGR}{}{}
\DeclareTextSymbol{\~}{LGR}{126}

\usepackage{ol2}

\makeatother

\begin{document}
\twocolumn[
\title{Ultra-Narrow Faraday Rotation Filter at the Rb D$_1$ Line }

\author{Joanna A. Zieli\'nska$^{1,2}$, Federica A. Beduini$^{1}$, Nicolas Godbout$^{3}$  and Morgan W. Mitchell$^{1,4,*}$}

\address{
$^1$ICFO-Institut de Ciencies Fotoniques, Mediterranean Technology Park, E-08860 Castelldefels (Barcelona), Spain\\
$^2$Instytut Fizyki Do\'swiadczalnej,  Uniwersytet Warszawski, PL-00-681 Warszawa,  Poland \\
$^3$COPL, D\'epartement de G\'enie Physique, \'Ecole Polytechnique
de Montr\'eal, C.P. 6079, Succ. Centre-ville, Montr\'eal
(Qu\'ebec) H3C 3A7, Canada\\
$^4$ICREA, Passeig Llu\'is Companys, 23, E-08010 Barcelona, Spain \\
$^*$Corresponding author: morgan.mitchell@icfo.es
}

\begin{abstract}
We present a theoretical and experimental study of the ultra-narrow bandwidth Faraday anomalous dispersion optical filter (FADOF) operating at the rubidium D$_1$ line (\SI{795}{nm}). 
This atomic line gives better performance than other lines for the main FADOF figures of merit, e.g. simultaneously $71\%$ transmission, \SI{445}{MHz} bandwidth and \SI{1.2}{GHz} equivalent noise bandwidth.
\end{abstract}


]

\hyphenation{FADOF FADOFs}

{
Ultra-narrow bandwidth optical filters are key elements in laser remote
sensing (LIDAR), observational astronomy, free-space communications and quantum optics.
Relative to conventional interference filters, 
{FADOFs} 
offer high background-rejection, mechanical robustness,
imaging capability and high transmission.  FADOFs have been developed for
several alkali atom resonances
 -- Cs D$_2$ \mycite{MendersOL1991} and $6S_{1/2}\rightarrow7P_{3/2}$  \mycite{Cessium} lines, Rb D$_2$ line
\mycite{RbD2report,RbD2}, K  (three lines)  \mycite{KD1},
Na D lines \mycite{Sodium}, and for Ca \mycite{Ca}.

We demonstrate
a FADOF on the D$_1$ line of Rb (wavelength \SI{795}{nm}).
This line, efficiently detected with Si detectors, accessible with a 
variety of laser technologies,
and with large hyperfine splittings, is a favorite for coherent and quantum
optics with warm atomic vapors.  Applications include electromagnetically-induced
transparency \mycite{EIT4WM}, stopped light \mycite{stopped}, optical
magnetometry \mycite{EITmag,WolfgrammPRL2010},  laser
oscillators \mycite{RbLaser}, polarization squeezing
\mycite{PolSqueezedLvovsky,PolSqueezedGrangier}, quantum
memory \mycite{memory} and high-coherence heralded single photons
\mycite{CereOL2009,WolfgrammPRL2011}.

Here we show that the Rb D$_1$ line provides superior FADOF performance.  
We demonstrate a FADOF surpassing other atoms and other Rb transitions for key figures of merit,
including peak transmission $T_{\rm max}$, transmission bandwidth and  equivalent noise bandwidth
${\rm ENBW}={T^{-1}_{\rm max}}\int T(\nu)d\nu$, where $T(\nu)$ is the filter transmission versus frequency $\nu$ \mycite{RbD2}.

}

\begin{figure}[htb]
 \centerline{\includegraphics[width=7.5cm]{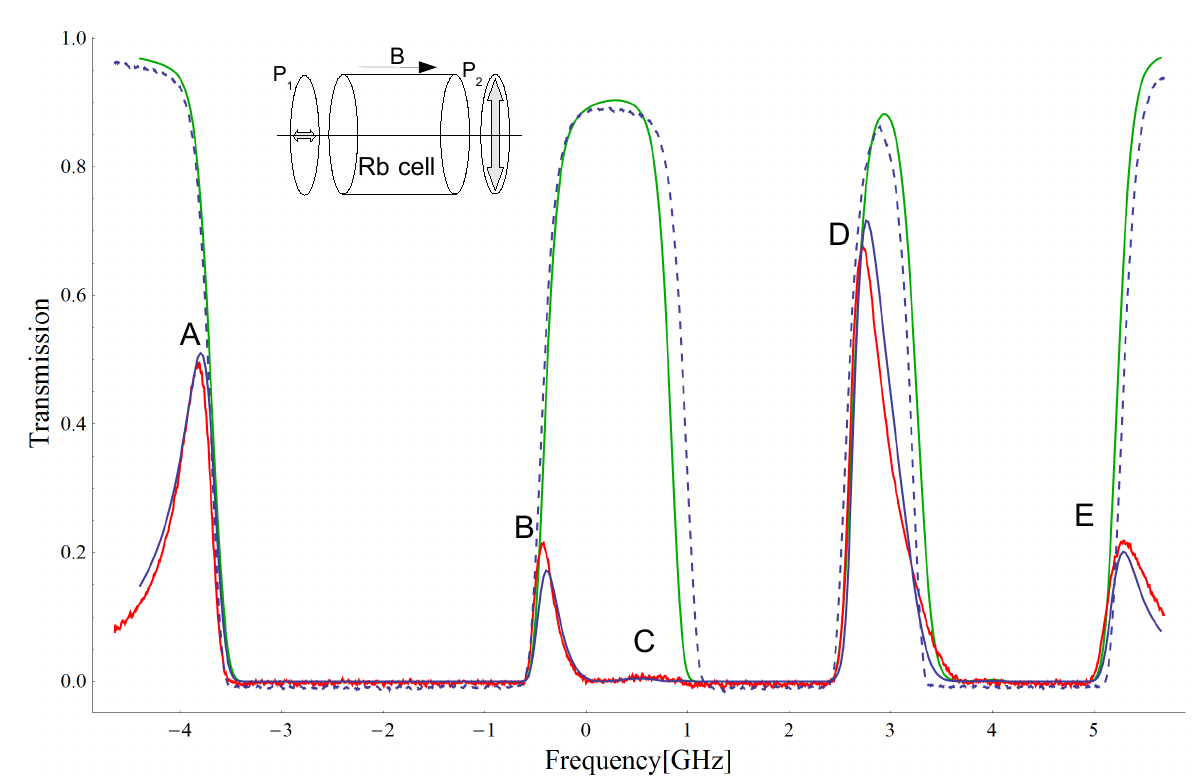}}
\caption{\label{fig:Theoretical-absorption-spectrum}
Absorption spectrum for Rb D$_{1}$: theory (green), experiment (black dashed).  FADOF transmission spectrum: theory (blue), experiment (red). Conditions:
$T=\SI{365}{K}$, $B=\SI{4.5}{mT}$. Inset - FADOF
setup.}
\end{figure}

{
A FADOF, shown schematically in Figure \ref{fig:Theoretical-absorption-spectrum} (inset), consists of an atomic vapor cell between two crossed polarizers.   A
homogeneous magnetic field along the propagation direction induces
circular birefringence in the vapor. The crossed polarizers
block transmission away from the absorption line, while the absorption itself
blocks resonant light.  Nevertheless, Faraday rotation just outside the Doppler
profile can give high transmission for a narrow range of frequencies.
FADOF is simple and robust, but performance depends critically on optical
properties of the atomic vapor.  We model these with a first-principles
calculation and find excellent agreement with
experiment, as shown in Figure \ref{fig:Theoretical-absorption-spectrum}.
}

\bigskip{}

{
\section*{Experiment}

\begin{figure}[htb]
 \centerline{\includegraphics[width=7.5cm]{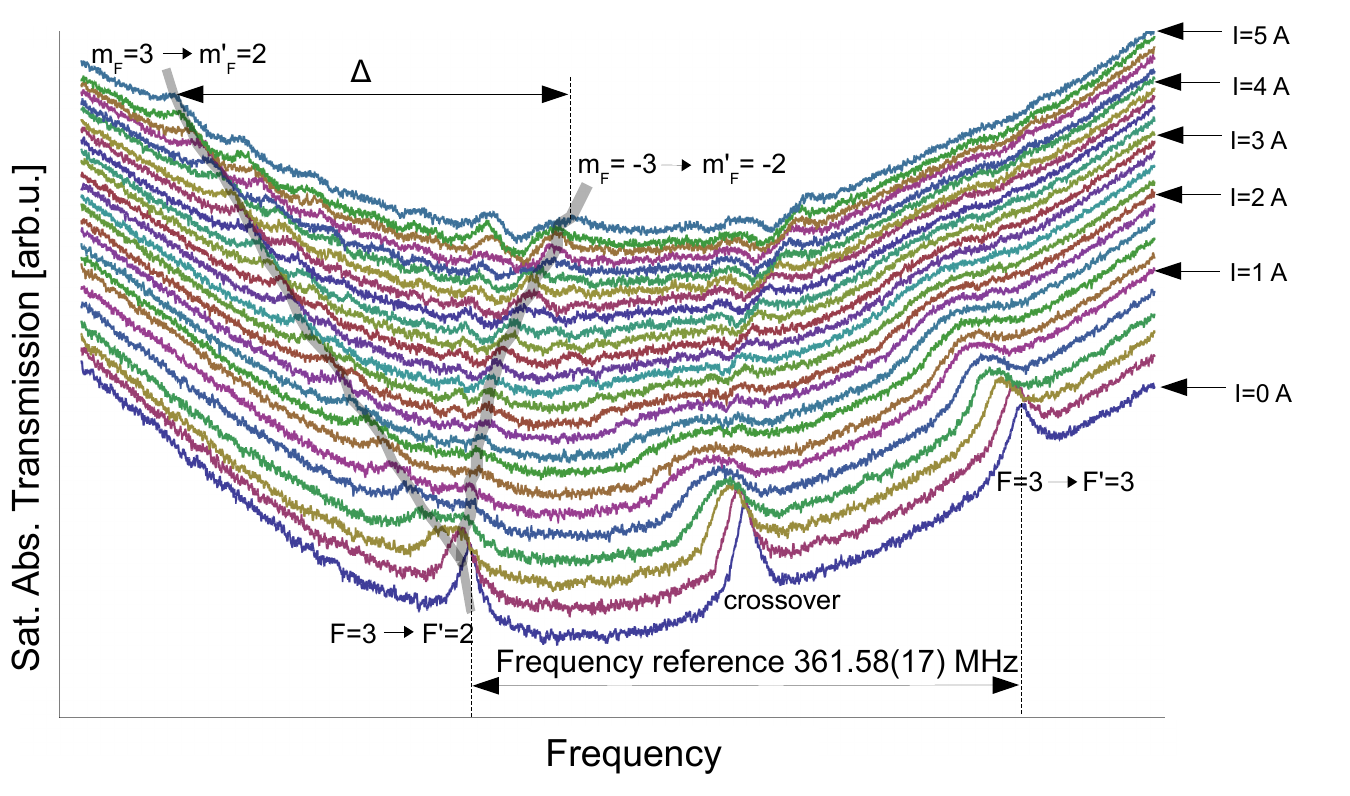}}
\caption{\label{fig:sas} Saturated absorption spectra of the Doppler-broadened line ($^{85}$Rb transitions from the $F=3$ ground state) for different coil currents. Splitting $\Delta$ between 
Zeeman-split lines provides a precise measure of the field.}
\end{figure}

We use a rubidium cell, with anti-reflection 
coated windows, a \SI{10}{cm}  internal path,
no buffer gas or wall coatings and natural abundance
rubidium (Technical Glass, Inc.).  An oven and
solenoidal field coil allow the cell to be maintained at constant
temperature in a uniform axial field.  Probe
light at \SI{795}{nm} from a Littrow-configuration external-cavity diode laser
(Toptica DL100) is spatially-filtered with a fiber and prepared in an
adjustable linear polarization using quarter- and half-waveplates.
After the cell, the beam is polarization-filtered with a  linear polarizer
(colorPol VIS IR) and detected with a low-noise amplified photodiode.}

Saturated absorption spectroscopy was used to calibrate field strength
produced by the solenoid current, as shown in
Figure \ref{fig:sas}.  Labeling the eigenstates by the total
angular momentum quantum numbers $(F,m_F)$, the 
 zero-field splitting between the $(3,m) \rightarrow (2,m)'$ and
 $(3,m) \rightarrow (3,m)'$ transitions of $^{85}$Rb provides a precise reference for 
the frequency scale, while the splitting {$\Delta$} between $(3,3) \rightarrow (2,2)'$
and $(3,-3) \rightarrow (2,-2)'$
is maximally sensitive to magnetic field.  Comparing the observed 
splittings to the first-principles
model, we find the field with  $\pm 0.6$ \% accuracy. 
Absorption spectra, acquired at the same time as the FADOF spectra, are used 
to calibrate the frequency axis, which otherwise would be non-linear due to 
piezo hysteresis.  These same spectra indicate the temperature with an 
uncertainty of $\pm$~\SI{2}{K}.  

{
The FADOF transmission was measured under a variety of temperature and field
conditions in the range T = 355--\SI{380}{K} and B = 3.3--\SI{6.0}{mT}.   Typical results, taken 
at \SI{4.5}{mT} and \SI{365}{K}, are shown in
Figure \ref{fig:Theoretical-absorption-spectrum} and show an agreement with
the model at the few-percent level.  This field/temperature combination gives a single
dominant peak with transmission of { 70\% at +\SI{2.8}{GHz} relative to line center with FWHM of \SI{445}{MHz}.  Other peaks, of transmission
 49\%, 17\% and 20\%} are also present, giving  an ENBW of
\SI{1.2}{GHz}, considerably better than reported for other atoms.  As shown in Table
\ref{tab:comparison}, the Rb D$_1$ FADOF also achieves narrower bandwidths
than other transitions and can achieve transmission up to 92\% consistent with ultra-low
ENBW.

{
\begin{table}
\centering
\caption{\label{tab:comparison}
Comparison of reported FADOF transmission for different atoms and wavelengths.  T$_{\rm max}$: peak transmission.   B$_{\rm T}$: full-width at half-maximum bandwidth of main transmission peak. B$_{\rm N}$: Equivalent-noise bandwidth (ENBW).  - value not reported.  Reference   \cite{KD1} also shows a K \SI{770}{nm} FADOF curve very similar to K \SI{766}{nm}.    }

\begin{tabular}{cccccc}
 &  &  & \tabularnewline
\hline
Atom& $\lambda$[nm]  & Ref.  & T$_{\rm max}$  &B$_{\rm T}$[GHz]&B$_{\rm N}$[GHz]\tabularnewline
\hline
K & 405 &\mycite{KD1}  & 0.93  & 1.2 & 6\tabularnewline
Ca &423 & \mycite{Ca}  & 0.55  & 1.5 & - \tabularnewline
Cs &455 &\mycite{Cessium}    & 0.96  & 0.9 & 3.3 \tabularnewline
Na &589 &\mycite{Sodium}  & 0.85  & 1.9& 5.1  \tabularnewline
Na &590 &\mycite{Sodium}   & 0.37  & 10.5 & 8.3 \tabularnewline
K &766 &\mycite{KD1} & 0.96  & 0.9 & 5   \tabularnewline
Rb &780 & \mycite{RbD2}   & 0.93  & 1.3& 4.7 \tabularnewline
Cs &852 &\mycite{MendersOL1991}   & 0.90  & 0.6 & - \tabularnewline
\hline
T[K] & B[mT]  & &  &   \tabularnewline
\hline
353 & 18.0   & & 0.92 & 0.48 & 2.1\tabularnewline
378 & 5.9  & & 0.91 & 1.10 & 2.7  \tabularnewline
365 & 4.5   & & 0.71 & 0.45& 1.2\tabularnewline
345 & 2.0  & & 0.04 & 0.32 & 0.8 \tabularnewline
\end{tabular}

\end{table}
}

The superior performance of the Rb D$_1$ line appears to be a fortunate accident of the hyperfine splittings.  For either pure $^{85}$Rb or pure $^{87}$Rb, the FADOF transmission at these field strengths shows four peaks, with the strongest ones at the extremes of the spectrum and with long tails.  The strong $^{87}$Rb peaks are visible as peak A and E of Figure \ref{fig:Theoretical-absorption-spectrum}.  The strong $^{85}$Rb peaks include one at \SI{-2.5}{GHz}, completely obscured by the $^{87}$Rb absorption, and the peak D.  The long tail of peak D is blocked by absorption from the $^{87}$Rb $1\rightarrow1$ transition, improving ENBW. 
}

\section*{Conclusions}

We have demonstrated a Faraday anomalous dispersion optical filter (FADOF) operating at the rubidium D1 line.  The filter gives high transmission and narrow bandwidth, with typical numbers being  0.5 GHz bandwidth, maximum transmission of 0.7 and equivalent-noise bandwidth of \SI{1.2}{GHz}, surpassing similar filters using other atoms and/or other optical transitions.  The spectrum can be optimized for different figures of merit by adjusting the temperature and magnetic field conditions of the atomic vapor.  A theoretical model shows excellent agreement with experimental results.  The simplicity and high noise-rejection may make the new filter attractive for LIDAR and free-space communications and introduces FADOF for the D$_1$ line, widely preferred for coherent and quantum optical applications.  

\section*{Acknowledgements}

We thank 
Y. A. de Icaza Astiz for discussions and assistance.
This work was supported by the Spanish Ministry of Science and
Innovation under projects FIS2008-01051, FIS2011-27806, 
and the Consolider-Ingenio 2010 Project ``Quantum
Optical Information Technologies.''

{
\section*{Appendix: FADOF spectra calculation}

For completeness, we present the full theoretical model of the Rb D$_1$ FADOF.  
A Mathematica notebook to perform the calculations is available with this document at \href{http://arxiv.org/find/all/1/all:+AND+mitchell+AND+godbout+AND+zielinska+beduini/0/1/0/all/0/1}{http://arxiv.org/}.

%

\newcommand{\Zeeman}{{\rm Ze}}
\newcommand{\hH}{\hat{H}}
\newcommand{\hI}{\hat{I}}
\newcommand{\hJ}{\hat{J}}
\newcommand{\hD}{\hat{D}}
\newcommand{\hQ}{\hat{Q}}

The Rb D transitions connect the $5^{2}S_{J=1/2}$ ground states to the $5^{2}P_{J=1/2}$ (D$_1$) and $5^{2}P_{J=3/2}$ (D$_2$) excited states.  Within each manifold,
the Zeeman/hyperfine structure is determined by the Hamiltonian
\begin{eqnarray}
\label{eq:Htot}
\hH &=& E_{\rm F} + \hH_{\Zeeman} + \hH_{\rm HF} \\
\hH_{\Zeeman}&=&-g_{J}\mu_{B}(\boldsymbol{B}\cdot\boldsymbol{\mathrm{\hat{J})}}\otimes\mathbb{\mathrm{\mathbb{I}}}_{2I+1}
\nonumber \\ & & -g_{I}\mu_{B}\mathbb{\mathrm{\mathbb{I}}}_{2J+1}\otimes(\boldsymbol{B}\cdot\mathrm{\mathbf{\hat{I}})}\\
\hH_{\rm HF}&=& A_{\rm HF} \hD_{\rm HF} + B_{\rm HF} \hQ_{\rm HF} \\
\hD_{\rm HF}&=& \boldsymbol{\mathrm{\hat{J}\circ}}\hat{\mathbf{\boldsymbol{I}}}
  \\
\hQ_{\rm HF}&=&\left\{ \begin{array}{ll} \frac{3(\boldsymbol{\mathrm{\hat{J}\circ}}\hat{\mathbf{\boldsymbol{I}}})^2+\frac{3}{2}\boldsymbol{\mathrm{\hat{J}\circ}}\hat{\mathbf{\boldsymbol{I}}}-I(I+1)J(J+1)}{2I(2I-1)J(2J-1)} & J>1/2 \\ 0 & {\rm otherwise} \end{array} \right.
\end{eqnarray}
where $E_{\rm F}$ is the level energy including the fine structure contribution,
$A_{\rm HF}$ and $B_{\rm HF}$ are magnetic dipole and quadrupole
energies, $\bf \hJ$ and $\bf \hI$ are vector operators representing the total electronic and nuclear
angular momenta, respectively,   $\boldsymbol{\mathrm{\hat{J}\circ}}\hat{\mathbf{\boldsymbol{I}}}\equiv \hJ_{x}\otimes \hI_{x}+\hJ_{y}\otimes \hI_{y}+\hJ_{z}\otimes \hI_{z}$, $g_{J}$ and $g_{I}$ are total electronic and nuclear gyromagnetic factors, respectively,
$\mathbb{\mathrm{\mathbb{I}}}_{n}$ is the $n$-dimensional identity matrix
and $\mu_{B}$ is the Bohr magneton.  All constants are taken from Steck
\mycite{SteckRb85,SteckRb87}.

The transition electric dipole operator $\bf \hD$, with circular components
$\hD_q$, $q = 0, \pm 1$,  is given by {
\begin{eqnarray}
D^{J,m_J,J',m_J'}_{q}
& =&\left\langle J||\hD||J'\right\rangle (-1)^{J'-1+m_{J}}
\nonumber \\
& & \times \sqrt{2J+1}
\left(\begin{array}{ccc}
J' & 1 & J\\
m'_{J} & q & m_{J}
\end{array}\right)
\end{eqnarray}
where last expression $(:::)$ is the Wigner 3-j symbol. The reduced dipole moment $\left\langle J||\hD||J'\right\rangle$ for each isotope is given by Steck \mycite{SteckRb85,SteckRb87}.

%
%

For  $^{85}$Rb ($I=5/2$) or $^{87}$Rb ($I=3/2$), and for the
ground and excited state manifolds,
the Hamiltonian of Eq. (\ref{eq:Htot}) is diagonalized to find transition frequencies $\omega_{ba}\equiv \omega_b-\omega_a$ and their corresponding dipole matrix elements $D^{ab}_q = \left< a\right| \hD_q \left|b\right> $, between all pairs of ground and excited states $\left|a\right>,\left|b\right>$.  The (linear) electric susceptibility tensor $\boldsymbol{\chi}$ is diagonal in the circular ($q=\pm 1$)  basis, and we compute its diagonal entries (in S.I. units) as}
\newcommand{\iso}{{Z}}
\begin{eqnarray}
\chi_{qq}(\omega)&=& \sum_{\iso\in\{85,87\}} \frac{N_{\iso}(T)}{\epsilon_{0}\hbar} \sum_{a,b} |D_{q}^{ab}|^{2}(\rho_{bb}-\rho_{aa}) \nonumber \\
& & \times V(\sigma,\frac{\Gamma}{2},\omega_{ba}-\omega)
\end{eqnarray}
where $N_{\iso}(T)$ is the atomic number density for the isotope $\iso$, $T$ is the temperature, $\varepsilon_0$ is the vacuum permittivity, $\rho \propto \exp[-H/(k_{\rm B} T)]$ is the atomic density matrix and $k_{\rm B}$ is the Boltzmann constant. The Voigt profile is
\begin{eqnarray}
V(\sigma,\frac{\Gamma}{2},\omega_{ba}-\omega)&=&i \sqrt{\frac{\pi}{2\sigma^2} }e^{x^2}\textrm{Erfc}(x),
\end{eqnarray}
where  $\rm Erfc$ is the complementary error function,
$x \equiv [{\frac{\Gamma}{2}+i(\omega_{ba}-\omega)}]/({\sqrt{2}\sigma})$,
 and
$\sigma=\omega_{ba}\sqrt{{k_{\rm B}T}/({M_\iso c^{2}})}$. Atomic masses $M_\iso$ and natural linewidths $\Gamma$ are given in \mycite{SteckRb85,SteckRb87}.

%
%
%
%

Atom number  density is calculated as $N_{\iso}(T) = c_{\iso}
N_{\iso}^{(\rm pure)}(T)$, where $c_{\iso}$ are the relative abundances,
 the single-isotope densities are
$N_{\iso}^{(\rm pure)}(T) = P_{\iso}(T)/(RT)$ where $R$ is the
ideal-gas constant and the single-isotope vapor pressures $P_{\iso}(T)$
are from Steck \mycite{SteckRb85,SteckRb87}.  We assume natural abundance  $c_{{85}} = 0.7217$ and  $c_{{87}} = 0.2783$.

%
%
%
%

The transmission of the filter is
\begin{equation}
T = | E_{\rm det}^\dagger\cdot M_{\rm cell} \cdot E_{\rm in} |^2,
\end{equation}
where $E_{\rm in} = (1,1)^T/\sqrt{2}$ is the Jones vector, in the circular left/right basis,
of the linearly polarized input,
$M_{\rm cell} = {\rm diag}(\exp[i n_L k L], \exp[i n_R k L])$ is the transmission matrix of the
cell, $n_{q} = \sqrt{1 + \chi_{qq}}$,  $E_{\rm det} = (1,-1)/\sqrt{2}$ is the Jones vector of the detected polarization, $k=\omega_{\rm laser}/c$ and $L$ is the cell internal path length.

%
%
%
%
%
%
%
%
%
%
%
%
%
%

\bigskip{}
}

\bibliographystyle{ol}
\bibliography{FADOF4}

\end{document}